\documentclass{article}

\usepackage{amsfonts}

\title{Classical and Quantized Affine Models of Structured Media}

\author{ Jan J. S\l awianowski\\
Institute of Fundamental Technological Research,\\
Polish Academy of Sciences,\\
\'{S}wi\c{e}tokrzyska 21, 00-049 Warsaw, Poland\\
e-mail: jslawian@ippt.gov.pl}

\begin{document}

\maketitle

\begin{abstract}
Having in view some applications in nanophysics, in particular in
nanophysics of materials, we develop new dynamical models of
structured bodies with affine internal degrees of freedom. In
particular, we construct some models where not only kinematics but
also dynamics of systems of affine bodies is affinely invariant.
Quantization schemes are developed. This is necessary in the range
of physical phenomena we are interested in.
\end{abstract}

\section*{Introduction}

The idea of microstructure is a rather old one and goes back to
brothers Cosserat who formulated the theory of continuum
consisting of infinitesimal gyroscopes \cite{Coss}. Eringen
modified such a model by introducing homogeneous deformations as
additional microstructural modes \cite{Erin}. These theories were
rather phenomenological and often motivated by some kind of
mathematical (more precisely, differential-geometric) aesthetics.
There was also mechanical motivation based on theory of granular
media and continuous limit of the dynamics of molecular crystals.

Recently the interest in mechanics of structured media becomes
more and more intensive in connection with nano-structures,
various supramolecular stru\-ctures, and defect theory. There are
also some special problems like the dynamics of suspensions, gas
bubbles in fluids, and some very peculiar models like kinetic
media \cite{Cap89,Cap00,Cap-Mar03(ed),Cap-Mar03,Mar}. In all these
problems, in defect theory and in dynamics of fullerens, affine
model of structural elements is well-motivated both from the
physical and geometrical point of view. There are some interesting
problems, completely new ones in comparison with traditional
phenomenological mechanics of structured continua. First of all,
in dynamics of strongly interacting systems of objects in
nano-scale, on the classical level, the traditional scheme of
constrained motion based on the d'Alembert principle does not seem
reliable any longer. Instead, one should base the dynamics on some
effective collective models motivated by appropriate symmetry
demands. We believe, there are physical reasons to expect that the
dynamics of internal and collective affine modes should be also
affinely invariant. In traditional theories derived from the
d'Alembert principle, the dynamics of affine modes is invariant
only under the Euclidean group. Below we suggest some dynamical
models of a system of bodies with affine degrees of freedom, with
Hamiltonians invariant under the affine group. One can smoothly
include also some terms invariant only under the Euclidean group,
there is however a natural temptation to consider them as a merely
perturbation to the background affine dynamics. Some arguments
from the solid state physics, like the concept of effective mass
support the idea that in complicated problems of condensed matter
theory the "true" metric tensor is not necessarily a fundamental
geometric background of equations of motion, Lagrangians and
Hamiltonians (having in mind microphysical models, we assume
Hamiltonian, variational dynamical models, without dissipation on
the fundamental level).

There is also another important novelty in comparison with
traditional macroscopic models. Namely, in the nano-scale the
quantum background of the dynamics must be seriously taken into
account. Because of this we developed some quantization scheme.
But at the same time one uses the concepts of macroscopic origin.
like, e.g., deformation tensors, deformation invariants, etc. It
seems rather amazing that one formulates questions like: what is a
complex quantum probability amplitude for the deformation tensors
and deformation invariants to be found in some fixed range? What
are the corresponding quantum transition probabilities? There are
even more serious problems. Namely, in this very peculiar range of
phenomena one has to do with a very complicated convolution of
classical and quantum problems. In this, rather unexpected way,
the old problems from the realm of foundations of quanta
\cite{Mac}, like decoherence, wave functions reduction,
possibility of nonlinear quantum description, etc. revive as ones
motivated by quite practical, structural physics. One must
honestly say, there is more secrets than well-established facts
and answers here.

\section{Affine bodies and their systems. General concepts and basic quantities}

There is no place here for the very detailed geometric
description. A rather exhaustive treatment of the
differential-geometric background of our ideas was presented in
some earlier papers (cf. \cite{all04,all05} and references there)
and a more detailed treatment will soon appear. Here we base
mainly on the analytic description where the physical and material
spaces are simply identified with $\mathbb{R}^{n}$ by some choice
of orthonormal Cartesian coordinates. As far as possible we work
in a non-specified dimension $n$ and only at some final stages we
specify $n$ to its physical value $3$ or to $2$ and $1$, which
obviously are also physically interpretable. This "false"
generality is mathematically convenient and better reveals some
structural features of the model, hidden behind the particular
value $n=3$. Moreover, it suggests even some analytical solving
procedures.

So, the body we consider consists of elements performing the translational motion in the
physical space $\mathbb{R}^{n}$ (physically $n=3,2,1$) and also some internal motion. The
latter is simply a relative motion of microconstituents; the translational
$\mathbb{R}^{n}$-degrees of freedom are attributed to the centres of mass. But of course one
can also admit situations when additional degrees of freedom are essentially internal ones,
like, e.g., in spin media. And in any case the distinction between internal and relative
motion may be (although need not be) historical, based on some conventions or on our
laboratory abilities. From now on we do not discuss this problem and for brevity all
non-translational degrees of freedom will be referred to as internal ones.

In mechanics of structureless continua the configuration space may be identified with
Diff$(n,\mathbb{R})$, i.e., the group of (sufficiently smooth) diffeomorphisms of
$\mathbb{R}^{n}$ onto itself. Let us remember, according to our analytical conventions both
the physical and material space are simply identified with $\mathbb{R}^{n}$. Any
diffeomorphism $\varphi\in$ Diff$(n,\mathbb{R})$ establishes an interrelation between
Lagrangian (material) and Eulerian (physical, current) coordinates, respectively $a^{k}$ and
$x^{i}$,
\begin{equation}\label{1}
x^{i}=\varphi^{i}\left(\ldots,a^{k},\ldots\right).
\end{equation}
In this way the configuration space of continuous medium is identified with the
infinite-dimensional group Diff$(n,\mathbb{R})$. In the Arnold description of incompressible
ideal fluid \cite{Arn} this group is constrained to SDiff$(n,\mathbb{R})$, i.e., the subgroup
of volume-preserving diffeomorphisms. Euler equations are interpreted in terms of
right-invariant geodetic Hamiltonian systems on this group. This approach turned out to be at
least heuristically effective in hydrodynamical problems.

Of course, Diff$(n,\mathbb{R})$ would be completely non-effective and just meaningless as the
configuration space of internal degrees of freedom of structured bodies. Nevertheless, there
exist both physical and analytical reasons to concentrate on models with degrees of freedom
based on appropriately chosen groups and their homogeneous spaces. From the physical and
computational point of view it is clear that one must use some geometrically well-motivated
finite-dimensional subgroups $G\subset$ Diff$(n,\mathbb{R})$, just Lie groups acting in
$\mathbb{R}^{n}$. The most traditional pattern is $G=$ SO$(n,\mathbb{R})$, i.e., rigid body
model of internal degrees of freedom. It is not excluded, especially in quantized theory that
the non-connected configuration space $G=$ O$(n,\mathbb{R})$ with mirror-reflected
configurations may be also acceptable. Other natural model with more degrees of freedom is
that of affinely-rigid, i.e., homogeneously deformable, body; $G=$ GL$^{+}(n,\mathbb{R})$ or
perhaps, as above, the total non-connected linear group $G=$ GL$(n,\mathbb{R})$, with
reflections admitted. One can also think about L$(n,\mathbb{R})$, the algebra of all $n\times
n$ matrices as an admissible configuration space. Incidentally, there is a subtle difference
between admitting mirror-reflected configurations of metrically- and affinely-rigid bodies.
Namely, GL$^{+}(n,\mathbb{R})$ and its mirror-reflected coset (not a subgroup!)
GL$^{-}(n,\mathbb{R})$ in GL$(n,\mathbb{R})$ infinitesimally approach each other in
L$(n,\mathbb{R})$, being separated only by the $(n^{2}-1)$-dimensional subset of singular
matrices in L$(n,\mathbb{R})$. On the other hand, SO$(n,\mathbb{R})$ and the set of improper
rotations O$(n,\mathbb{R})\backslash$ SO$(n,\mathbb{R})$ (the complement of SO$(n,\mathbb{R})$
in O$(n,\mathbb{R})$) are so-to-speak finitely separated in L$(n,\mathbb{R})$.

There are also natural models placed between gyroscopic and affine degrees of freedom; one
deals then with a constrained affinely-rigid body. Let us mention incompressible (isochoric)
body when $G=$ SL$(n,\mathbb{R})$, the special linear group. Just as previously, one can also
think about admitting mirror reflections. Then $G=$ UL$(n,\mathbb{R})=\left\{\varphi\in{\rm
GL}(n,\mathbb{R}):\left|\det\varphi\right|=1\right\}$, the unimodular group. It is a union of
SL$(n,\mathbb{R})$ and of its coset in GL$(n,\mathbb{R})$ consisting of matrices with the
minus-one-determinant. Obviously, just as in the case of metrically-rigid body, the manifolds
of proper and improper (orientation-changing) isochoric mappings are finitely-separated in
L$(n,\mathbb{R})$. In a sense an opposite model (less realistic) is that of shape-preserving
affinely-rigid body when, roughly speaking, internal configurations are built of dilatations
and rotations, $G=\mathbb{R}^{+}$ SO$(n,\mathbb{R})$; $\mathbb{R}^{+}$ denoting the
multiplicative group of positive real numbers. Again, admitting orientation-preserving
mappings we have $G=\mathbb{R}^{+}$ O$(n,\mathbb{R})$. And here also the two connected
components approach each other infinitely close. The closure of $G$ contains the null matrix,
therefore, it is not a subset of GL$(n,\mathbb{R})$.

All these subgroups of Diff$(n,\mathbb{R})$ are contained in GL$(n,\mathbb{R})$; they consist
of linear mappings. These mappings describe configurations in such a way that the material
point with Lagrangian coordinates $a^{k}$ occupies the spatial position with Euler coordinates
$y^{i}$,
\begin{equation}\label{2}
y^{i}=x^{i}+\varphi^{i}{}_{k}a^{k},
\end{equation}
where $x^{i}$ are spatial coordinates of the centre of mass. The very geometry suggests also
some groups of non-affine transformations, e.g., the projective group GPr$(n,\mathbb{R})$,
conformal group Co$(n,\mathbb{R})$. One can realize physical situations where the relevant
collective modes of internal motion are described just by these groups. It is also
non-excluded that the complex group GL$(n,\mathbb{C})$ or its unitary subgroup U$(n)\subset$
GL$(n,\mathbb{C})$ \cite{Westph} may be useful. In any case, it happens quite often in physics
that the complexification leads to quite amazing, unexpected new results.

Just as in our earlier papers we concentrate here on the affine model, when $G=$
GL$(n,\mathbb{R})$. Incidentally, the projective model may be in some sense reduced to it,
because GPr$(n,\mathbb{R})$ is isomorphic with SL$(n+1,\mathbb{R})$. In any case, affine modes
seem to be dominant for small structure entities like molecules, microdefects, some
supramolecular clusters, fullerens, etc.

The configuration space of a single structural elements is identified with
\begin{equation}\label{3}
Q=\mathbb{R}^{n}\times G=Q_{\rm tr}\times Q_{\rm int},
\end{equation}
in particular, for elements with affine modes of deformation:
\begin{equation}\label{4}
Q=\mathbb{R}^{n}\times {\rm GL}(n,\mathbb{R}).
\end{equation}
Usually, especially in classical (non-quantized) problems, GL$(n,\mathbb{R})$ is replaced by
GL$^{+}(n,\mathbb{R})$. The labels "tr" and "int" refer obviously to translational and
internal degrees of freedom. The total body (medium) consists of $N$ elements. Its
configuration space is obviously given by the Cartesian product
\[
Q^{N}=Q^{N}_{\rm tr}\times Q^{N}_{\rm int}\simeq \mathbb{R}^{nN}\times G^{N};
\]
in our treatment GL$(n,\mathbb{R})$ or GL$^{+}(n,\mathbb{R})$ substituted for $G$. Therefore,
the configuration is an array:
\begin{equation}\label{5}
q=\left(x_{1},\ldots,x_{N};\varphi_{1},\ldots,\varphi_{N}\right),
\end{equation}
where $x_{A}\in\mathbb{R}^{n}$, $\varphi_{A}\in{\rm GL}(n,\mathbb{R})$, and
$A=\overline{1,N}$.

For a single structure element the summation of usual kinetic energies of its constituents
gives in virtue of (\ref{2}), the usual d'Alembert form
\begin{equation}\label{6}
T^{\rm d'A}=T^{\rm d'A}_{\rm tr}+T^{\rm d'A}_{\rm int}=\frac{M}{2}{\rm Tr}\left(vv^{T}\right)
+\frac{1}{2}{\rm Tr}\left(\xi J\xi^{T}\right),
\end{equation}
where $v\in \mathbb{R}^{n}$, $\xi\in$ L$(n,\mathbb{R})$ denote respectively the translational
and internal velocities:
\begin{equation}\label{7}
v^{i}=\frac{dx^{i}}{dt},\qquad \xi^{i}{}_{k}=\frac{d\varphi^{i}{}_{k}}{dt},
\end{equation}
and $M$, $J$ are constant inertial characteristics. More precisely, $M$ is the total mass of
the element and the symmetric positively definite matrix $J$ is its modified inertial tensor,
i.e., second-order moment of the mass distribution with respect to co-moving (Lagrange)
coordinates,
\begin{equation}\label{8}
M=\sum_{p}\mu_{p},\qquad J^{kl}=\sum_{p}\mu_{p}a^{k}_{p}a^{l}_{p}.
\end{equation}
Summation is performed over constituents ("atoms") of the element ("mole\-cule"); $\mu_{p}$ is
the mass of the $p$-th constituent. Sometimes it is convenient to use the symbol of
integration with respect to the mass distribution measure $\mu$:
\begin{equation}\label{9}
J^{kl}=\int a^{k}a^{l}d\mu(a).
\end{equation}

\noindent{\bf Remark:} the kinetic energy (\ref{6}) is spatially
isotropic, i.e., invariant under the transformations $L_{A}$ below
(\ref{13}) with $A$ restricted to the orthogonal group
O$(n,\mathbb{R})$ (spatial rotations). So are its both terms
separately. The material rotations $R_{A}$ in (\ref{13}) preserve
$T_{\rm tr}$ trivially, but in general $T_{\rm int}$ is
non-invariant under the right-acting O$(n,\mathbb{R})$. However,
it is invariant under the right actions of O$(n,J)$, the subgroup
of GL$(n,\mathbb{R})$ preserving $J$. This reduces to the
O$(n,\mathbb{R})$-invariance, when $J$ is isotropic, i.e.,
$J=I{\rm Id}_{n}$; $I$ is a positive constant of internal inertia
and Id$_{n}$ is the $n\times n$ identity matrix. Then
\begin{equation}\label{6a}
T^{\rm d'A}_{\rm int}=\frac{I}{2}{\rm Tr}\left(\xi\xi^{T}\right).
\end{equation}

The total kinetic energy of the body is given by
\begin{equation}\label{10}
T^{\rm d'A}=\sum^{N}_{A=1}T^{\rm d'A}_{A}=\frac{1}{2}\sum^{N}_{A=1}M_{A}{\rm
Tr}\left(v_{A}v_{A}^{T}\right)+\frac{1}{2}\sum^{N}_{A=1}{\rm
Tr}\left(\xi_{A}J_{A}\xi_{A}^{T}\right).
\end{equation}
Assuming that the body consists of identical structure elements we have that $M_{A}=M$,
$J_{A}=J$, $A=\overline{1,N}$.

Let us again concentrate on a single element. Its Green and Cauchy deformation tensors are
respectively denoted as
\begin{equation}\label{11}
G[\varphi]=\varphi^{T}\varphi=G[\varphi]^{T},\qquad
C[\varphi]=\varphi^{-1T}\varphi^{-1}=C[\varphi]^{T},
\end{equation}
similarly, for their contravariant inverses we write
\begin{equation}\label{12}
\widetilde{G}[\varphi]=\varphi^{-1}\varphi^{-1T},\qquad
\widetilde{C}[\varphi]=\varphi\varphi^{T}.
\end{equation}
Spatial and material transformations are respectively given by left and right regular
translations:
\begin{equation}\label{13}
\varphi\mapsto L_{A}(\varphi)=A\varphi,\qquad \varphi\mapsto R_{A}(\varphi)=\varphi A
\end{equation}
for any fixed $A\in$ GL$(n,\mathbb{R})$. When $A\in$ O$(n,\mathbb{R})$, then obviously
\begin{equation}\label{14}
G[A\varphi]=G[\varphi],\qquad C[\varphi A]=C[\varphi],
\end{equation}
and for the general $A\in$ GL$(n,\mathbb{R})$
\begin{equation}\label{15}
G[\varphi A]=A^{T}G[\varphi]A,\qquad C[A\varphi]=A^{-1T}C[\varphi]A^{-1}.
\end{equation}
There is no concise formula for $G[A\varphi]$, $C[\varphi A]$ if $A$ is not orthogonal (does
not belong to O$(n,\mathbb{R})$).

Deformation invariants are scalar functions $f:$ GL$(n,\mathbb{R})\rightarrow\mathbb{R}$
invariant under (\ref{13}) for orthogonal translations
\begin{equation}\label{16}
f(A\varphi B)=f(\varphi)
\end{equation}
for any $A,B\in$ O$(n,\mathbb{R})$. There are $n$ basic invariants through which all other
ones may be expressed. Various choices are possible, e.g., the following frequently used
\begin{equation}\label{17}
\mathcal{K}_{a}[\varphi]={\rm Tr}\left(G[\varphi]^{a}\right)={\rm
Tr}\left(C[\varphi]^{-a}\right),\qquad a=\overline{1,n},
\end{equation}
eigenvalues $\lambda_{a}[\varphi]$ of $G[\varphi]$,
\begin{equation}\label{18}
\det\left(G[\varphi]-\lambda[\varphi]I_{n}\right)=0,
\end{equation}
or coefficients $I_{p}[\varphi]$ of the eigenequation
\begin{equation}\label{19}
\det\left(G[\varphi]-\lambda I_{n}\right)=\sum^{n}_{k=0}(-1)^{k}I_{n-k}[\varphi]\lambda^{k};
\end{equation}
obviously, $I_{0}=1$ is standard. Geometrically speaking, deformation invariants are functions
on the manifold of double cosets
\[
{\rm Inv}:={\rm O}(n,\mathbb{R})\backslash{\rm GL}(n,\mathbb{R})/{\rm O}(n,\mathbb{R}).
\]
Deformation invariants are used when constructing potential energy models for a single affine
body. When dealing with the system of such bodies we need some basic scalars assigned to pairs
of internal configurations. In analogy to Green and Cauchy deformation tensors for any pair
$\psi,\varphi\in$ GL$(n,\mathbb{R})$ we define the quantities
\begin{equation}\label{20}
G[\psi,\varphi]:=\psi^{T}\varphi,\qquad C[\psi,\varphi]:=\varphi^{-1T}\psi^{-1}.
\end{equation}
Obviously,
\[
G[\psi,\psi]=G[\psi],\qquad C[\psi,\psi]=C[\psi],
\]
i.e., the above mutual deformation tensors reduce then to the usual ones.

But one can also define another mutual quantities, namely,
\begin{equation}\label{21}
\Gamma[\psi,\varphi]:=\psi^{-1}\varphi,\qquad
\Sigma[\psi,\varphi]:=\varphi\psi^{-1}.
\end{equation}
For orthogonal matrices they reduce to the previous ones,
\begin{equation}\label{22}
\psi,\varphi\in{\rm
O}(n,\mathbb{R})\Rightarrow\Gamma[\psi,\varphi]=G[\psi,\varphi],\quad
\Sigma[\psi,\varphi]=C[\psi,\varphi].
\end{equation}
Obviously, $\Gamma[\psi,\varphi]$, $\Sigma[\psi,\varphi]$ are
exactly group-theoretical counterparts of the displacement vector
in translational degrees of freedom. Indeed, interpreting
$\mathbb{R}^{n}$ as an Abelian group under addition of vectors, we
immediately notice that the prescription (\ref{21}) in the
non-Abelian multiplicative matrix group GL$(n,\mathbb{R})$ has
exactly the same group meaning as $u=y-w$ in $\mathbb{R}^{n}$.

It is clear that for any $A\in$ O$(n,\mathbb{R})$
\begin{equation}\label{23}
G[A\psi,A\varphi]=G[\psi,\varphi],\qquad C[\psi A,\varphi
A]=C[\psi,\varphi],
\end{equation}
i.e., they are respectively invariant under spatial and material
isometries. For the general $A\in$ GL$(n,\mathbb{R})$ we have
\begin{equation}\label{24}
G[\psi A,\varphi A]=A^{T}G[\psi,\varphi]A,\qquad
C[A\psi,A\varphi]=A^{-1T}C[\psi,\varphi]A^{-1}.
\end{equation}
And just as previously there is no concise expression for
$G[A\psi,A\varphi]$, $C[\psi A,\varphi A]$ if $A$ is
non-orthogonal.

Transformation rules for $\Gamma$, $\Sigma$ have another form.
Namely, for any $A\in$ GL$(n,\mathbb{R})$ we have
\begin{eqnarray}
\Gamma[A\psi,A\varphi]=\Gamma[\psi,\varphi],&\quad&
\Sigma[A\psi,A\varphi]=A\Sigma[\psi,\varphi]A^{-1},\label{26}\\
\Gamma[\psi A,\varphi A]=A^{-1}\Gamma[\psi,\varphi]A,&\quad&
\Sigma[\psi A,\varphi A]=\Sigma[\psi,\varphi].\label{27}
\end{eqnarray}
Therefore, $\Gamma$ is invariant under spatial affine
transformations and suffers the inverse adjoint rule under
material affine transformations. And conversely, $\Sigma$
transforms according to the adjoint rule under spatial affine
mappings and is affinely invariant under material transformations.

The quantities $G[\psi,\varphi]$, $C[\psi,\varphi]$,
$\Gamma[\psi,\varphi]$, $\Sigma[\psi,\varphi]$, give rise to
scalars which may be used as arguments of the potential energy
terms. Typical scalars of this type are, in analogy to (\ref{17}),
given by
\begin{equation}\label{28}
\mathcal{K}_{a}[\psi,\varphi]={\rm
Tr}\left(G[\psi,\varphi]^{a}\right)={\rm
Tr}\left(C[\psi,\varphi]^{-a}\right),\qquad a=\overline{1,n}.
\end{equation}
Just as in the case of deformation invariants, these scalars are
invariant under spatial and materia rotations (left and right
regular translations of $\varphi$, $\psi$ by orthogonal matrices):
\begin{equation}\label{29}
\mathcal{K}_{a}[A\psi B,A\varphi
B]=\mathcal{K}_{a}[\psi,\varphi],\qquad A,B\in{\rm
O}(n,\mathbb{R}).
\end{equation}
In analogy to (\ref{18}), (\ref{19}) one can also use solutions of
the eigenequation for $G[\psi,\varphi]$ (or $C[\psi,\varphi]$), or
coefficients in the eigenequation as basic invariants. Another
kind of invariants is built of $\Gamma,\Sigma$-objects, e.g.,
\begin{equation}\label{30}
\mathcal{M}_{a}[\psi,\varphi]={\rm
Tr}\left(\Gamma[\psi,\varphi]^{a}\right)={\rm
Tr}\left(\Sigma[\psi,\varphi]^{a}\right)
\end{equation}
like in (\ref{17}), or, according to the $\lambda_{a}$-,
$I_{a}$-schemes like in (\ref{18}), (\ref{19}). These objects are
invariant under all affine spatial and material transformations,
i.e.,
\begin{equation}\label{31}
\mathcal{M}_{a}[A\psi B,A\varphi
B]=\mathcal{M}_{a}[\psi,\varphi],\qquad A,B\in{\rm
O}(n,\mathbb{R})
\end{equation}
for any $A,B\in$ GL$(n,\mathbb{R})$. These scalars measures of the
"distance" between internal configurations are affinely invariant.
Unlike this, the measures $\mathcal{K}_{a}$ are only orthogonally
invariant, so they are usual Euclidean distances.

Let us remind that in many problems it is convenient to use
deformation measures which vanish in the non-deformed state, e.g.,
Lagrange and Euler deformation tensors:
\begin{equation}\label{32}
E[\varphi]=\frac{1}{2}\left(G[\varphi]-I\right),\qquad
e[\varphi]=\frac{1}{2}\left(I-C[\varphi]\right).
\end{equation}
By analogy, it may be convenient to replace the mutual tensors
$G[\psi,\varphi]$, $C[\psi,\varphi]$, $\Gamma[\psi,\varphi]$,
$\Sigma[\psi,\varphi]$ by
\begin{eqnarray}
E[\psi,\varphi]=\frac{1}{2}\left(G[\varphi]-I\right),&\quad&
e[\psi,\varphi]=\frac{1}{2}\left(I-C[\varphi]\right),\label{33}\\
\gamma[\psi,\varphi]=\Gamma[\psi,\varphi]-I,&\quad&
\sigma[\psi,\varphi]=\Sigma[\psi,\varphi]-I.\label{33a}
\end{eqnarray}
Another possibility is to use the exponential representation of
matrices $G$, $C$, $\Gamma$, $\Sigma$. It may be also convenient
to use invariants built of them according to the schemes like
(\ref{17}), (\ref{30}) or (\ref{18}), (\ref{19}), etc. Obviously,
such invariants are functionally dependent on the previous ones.

Affine velocity, i.e., Eringen's "gyration" \cite{Erin}
respectively in the spatial and co-moving representations is given
by
\begin{equation}\label{34}
\Omega=\frac{d\varphi}{dt}\varphi^{-1},\qquad
\widehat{\Omega}=\varphi^{-1}\frac{d\varphi}{dt}.
\end{equation}
Spatial and material transformations (\ref{13}) act on the above
quantities as follows:
\begin{eqnarray}
L_{A}&:&\Omega\mapsto A\Omega A^{-1},\qquad \widehat{\Omega}\mapsto\widehat{\Omega},\label{35}\\
R_{A}&:&\Omega\mapsto \Omega,\qquad \widehat{\Omega}\mapsto
A^{-1}\widehat{\Omega}A.\label{36}
\end{eqnarray}
When the motion is metrically rigid, i.e., permanently
$\varphi\in$ O$(n,\mathbb{R})$, then
\begin{equation}\label{37}
\Omega=-\Omega^{T},\qquad \widehat{\Omega}=-\widehat{\Omega}^{T}
\end{equation}
and these skew-symmetric objets reduce to the usual angular
velocity, respectively in the spatial and co-moving
representations. In some formulas it is convenient to use also the
co-moving representation of translational velocity. It is given by
\begin{equation}\label{38}
\widehat{v}=\varphi^{-1}v.
\end{equation}

Gyroscopic constraints may be described in anholonomic terms
simply by stating that $\Omega$ is skew-symmetric (and so is
$\widehat{\Omega}$ then),
\begin{equation}\label{39}
\Omega+\Omega^{T}=0.
\end{equation}
By analogy, constraints of the purely deformative rotationless
motion may be defined by the demand of permanently symmetric
$\Omega$,
\begin{equation}\label{40}
\Omega-\Omega^{T}=0.
\end{equation}
Let us observe that the materially rotationless constraints
\begin{equation}\label{41}
\widehat{\Omega}-\widehat{\Omega}^{T}=0
\end{equation}
are non-equivalent to the above spatially rotationless ones.

There is an interesting novelty now. Namely, rotationless
constraints are essentially non-holonomic. The point is that the
subspace of symmetric matrices is not a commutator Lie subalgebra.

Canonical momenta, i.e., dual objects of velocities (linear
functions of them), $p$, $\pi$ are elements of $\mathbb{R}^{n}$,
L$(n,\mathbb{R})$ respectively, and their pairing with velocities
is given by
\begin{equation}\label{42}
\langle p,v\rangle=p^{T}v={\rm Tr}\left(pv^{T}\right),\qquad
\langle\pi,\xi\rangle={\rm Tr}\left(\pi\xi\right).
\end{equation}
The co-moving representation of $p$ is given by
\begin{equation}\label{43}
\widehat{p}=\varphi^{T}p,
\end{equation}
and obviously, with this convention
\begin{equation}\label{44}
\langle
p,v\rangle=p^{T}v=\widehat{p}^{T}\widehat{v}=\langle\widehat{p},\widehat{v}\rangle.
\end{equation}

It is convenient to introduce non-holonomic canonical momenta
conjugate to $\Omega$, $\widehat{\Omega}$, namely,
\begin{equation}\label{45}
\Sigma:=\varphi\pi,\qquad \widehat{\Sigma}:=\pi\varphi.
\end{equation}
Obviously,
\begin{equation}\label{46}
\langle\Sigma,\Omega\rangle={\rm Tr}\left(\Sigma\Omega\right)={\rm
Tr}\left(\widehat{\Sigma}\widehat{\Omega}\right)=\langle\widehat{\Sigma},\widehat{\Omega}\rangle.
\end{equation}
Transformation rules for $\Sigma$, $\widehat{\Sigma}$ are
identical with (\ref{35}), (\ref{36}):
\begin{eqnarray}
L_{A}&:&\Sigma\mapsto A\Sigma A^{-1},\qquad \widehat{\Sigma}\mapsto\widehat{\Sigma},\label{47}\\
R_{A}&:&\Sigma\mapsto \Sigma,\qquad \widehat{\Sigma}\mapsto
A^{-1}\widehat{\Sigma}A.\label{48}
\end{eqnarray}
The components of $\Sigma$, $\widehat{\Sigma}$ are respectively
Hamiltonian generators of spatial and material affine
transformations (\ref{13}). Their doubled skew-symmetric parts
\begin{equation}\label{49}
S=\Sigma-\Sigma^{T}=-S,\qquad
V=\widehat{\Sigma}-\widehat{\Sigma}^{T}=-V
\end{equation}
are, respectively, Hamiltonian generators of spatial and material
rotations of internal degrees of freedom.

The relationship between Hamiltonian quantities like $p$,
$\widehat{p}$, $\Sigma$, $\widehat{\Sigma}$ and kinematical ones
like $v$, $\widehat{v}$, $\Omega$, $\widehat{\Omega}$ may be
established only on the basis of some particular model, when
Lagrangian $L$ is fixed. In potential models $L=T-\mathcal{V}$
with the potential $\mathcal{V}$ depending only on the
configuration $(x,\varphi)\in \mathbb{R}^{n}\times$
GL$(n,\mathbb{R})$, Legendre transformation has the following
form:
\begin{equation}\label{50}
p_{i}=\frac{\partial L}{\partial v^{i}}=\frac{\partial T}{\partial
v^{i}},\qquad \pi^{A}{}_{i}=\frac{\partial L}{\partial
\xi^{i}{}_{A}}=\frac{\partial T}{\partial \xi^{i}{}_{A}}.
\end{equation}
Assuming the d'Alembert model of the kinetic energy (\ref{6}) we
obtain that
\begin{equation}\label{51}
p=Mv,\qquad \pi=J\xi^{T},
\end{equation}
and the corresponding expression for the kinetic Hamiltonian
\begin{equation}\label{52}
\mathcal{T}=\mathcal{T}_{\rm tr}+\mathcal{T}_{\rm
int}=\frac{1}{2M}{\rm Tr}\left(pp^{T}\right)+\frac{1}{2}{\rm
Tr}\left(\pi^{T}J^{-1}\pi\right).
\end{equation}
For Hamiltonian systems $H=\mathcal{T}+\mathcal{V}$ ($\mathcal{V}$
denoting the potential energy) equations of motion may be
effectively analyzed in terms of Poisson brackets and canonical
Hamilton equations,
\begin{equation}\label{53}
\frac{dF}{dt}=\left\{F,H\right\},
\end{equation}
where $F$ runs over some maximal system of functionally
independent phase-space functions.

\section{Dynamical models. Affine invariance problems.
Realistic questions, academic questions, and pure fantasy}

For the system of affine bodies Lagrangian has the form:
\begin{equation}
L=T-\mathcal{V},\label{54}
\end{equation}
where the kinetic energy is obtained by summation of individual kinetic energies like in
(\ref{10}),
\begin{equation}
T=\sum^{N}_{A=1}T_{A}=T_{\rm tr}+T_{\rm int}=\sum^{N}_{A=1}\left( T_{\rm tr}\right)
_{A}+\sum^{N}_{A=1}\left( T_{\rm int}\right) _{A}. \label{55}
\end{equation}
The potential energy in typical situations consists of two main terms, the one- and two-body
potentials,
\begin{equation}
\mathcal{V}=\mathcal{V}^{(1)}+\mathcal{V}^{(2)}.  \label{56}
\end{equation}
It is known that in realistic problems it is usually less then 10\% of energy that could be
assigned to three-body and higher multibody interactions. $\mathcal{V}^{(1)}$ is the sum of
terms depending on individual elements,
\begin{equation}
\mathcal{V}^{(1)}\left( \ldots ;x_{A},\varphi _{A};\ldots \right)
=\sum^{N}_{B=1}\mathcal{V}^{(1)}{}_{B}\left( x_{B},\varphi _{B}\right).  \label{57}
\end{equation}
The over-simplified models where $\mathcal{V}^{(1)}{}_{B}$ splits into the sum of
translational and internal parts,
\begin{equation}
\mathcal{V}^{(1)}{}_{B}\left( x_{B},\varphi _{B}\right) =\mathcal{V} ^{(1)}_{\rm
tr}{}_{B}\left( x_{B}\right) +\mathcal{V}^{(1)}_{\rm int}{}_{B} \left( \varphi _{B}\right),
\label{58}
\end{equation}
are not very realistic, nevertheless, they provide some so-to-speak zeroth-order
approximation. When the elements are identical, all $\mathcal{V}^{(1)}{}_{B}$ have the same
functional form.

The binary term has the usual form,
\begin{equation}
\mathcal{V}^{(2)}\left( \ldots ;x_{A},\varphi _{A};\ldots \right) =\frac{1}{2
}\sum^{N}_{K,L=1}\mathcal{V}^{(2)}{}_{KL}\left( x_{K},\varphi _{K};x_{L},\varphi _{L}\right).
\label{59}
\end{equation}
And again the simplest, although rather academic models are those with the separated
dependence of $\mathcal{V}^{(2)}$ on translational and internal variables,
\begin{equation}
\mathcal{V}^{(2)}{}_{KL}\left( x_{K},\varphi _{K};x_{L},\varphi _{L}\right) =
\mathcal{V}^{(2)}_{\rm tr}{}_{KL}\left( x_{K},x_{L}\right) +\mathcal{V} ^{(2)}_{\rm
int}{}_{KL}\left( \varphi _{K},\varphi _{L}\right). \label{60}
\end{equation}
Mutual interactions should be translationally invariant, i.e., $\mathcal{V}^{(2)}{}_{KL}$
depend on $x_{K}$, $x_{L}$ through $\overrightarrow{x_{K}x_{L}}=x_{L}-x_{K}$. Isotropy of the
physical space implies that the radius-vectors $x_{L}-x_{K}$ enter $\mathcal{V}^{(2)}{}_{KL}$
only through their lengths $\|x_{L}-x_{K}\|$. There is some more discussion concerning the
dependence of $\mathcal{V}^{(2)}{}_{KL}$ on internal degrees of freedom. Isotropy of the
physical space implies that $\mathcal{V}^{(2)}_{\rm int}{}_{KL}$ should depend on
$\varphi_{K}$, $\varphi_{L}$ only through the mutual tensors $G[\varphi_{K},\varphi_{L}]$,
$\Gamma[\varphi_{K},\varphi_{L}]$, thus,
\begin{equation}
\mathcal{V}^{(2)}{}_{KL}\left( x_{k},\varphi _{k};x_{L},\varphi _{L}\right) =f_{KL}\left(
\left\| x_{L}-x_{K}\right\| ,G\left[ \varphi _{K},\varphi _{L} \right] ,\Gamma \left[ \varphi
_{K},\varphi _{L}\right] \right) \label{61}
\end{equation}
and obviously for the body consisting of identical elements there is no dependence on $K$,
$L$; $f_{KL}=f$ for some fixed $f$. And if the dynamics is to be invariant also under
simultaneous material rotations, then at the same time, $\mathcal{V}^{(2)}{}_{KL}$ must depend
on internal configurations only through $C[\varphi_{K},\varphi_{L}]$,
$\Sigma[\varphi_{K},\varphi_{L}]$. But this means that $\mathcal{V}^{(2)}{}_{KL}$ is
algebraically built of the mutual invariants, e.g., chosen as
$\mathcal{K}_{a}[\varphi_{K},\varphi_{L}]$, $\mathcal{M}_{a}[\varphi_{K},\varphi_{L}]$,
\begin{equation}
\mathcal{V}^{(2)}{}_{KL}\left( x_{K},\varphi _{K};x_{L},\varphi _{L}\right) =f_{KL}\left(
\left\| x_{L}-x_{K}\right\| ,\mathcal{K}\left[ \varphi _{K},\varphi _{L}\right]
,\mathcal{M}\left[ \varphi _{K},\varphi _{L}\right] \right). \label{62}
\end{equation}
In the last formula, $\mathcal{K}$, $\mathcal{M}$ are abbreviations for the systems
$\mathcal{K}_{a}$, $\mathcal{M}_{a}$, $a=\overline{1,n}$.

In our model, geometry of degrees of freedom and kinematics is
ruled by the affine group. On the other hand, the dynamics is not
invariant either under spatial or material affine transformations
(\ref{13}). The spatial metric tensor and the inertial moment $J$
break the affine symmetry and restrict it to the Euclidean one in
the physical space and to O$(n,J)$ in the material space. What
concerns potential energy of mutual interactions (\ref{62}) it is
clear that the vector norm $\|x_{L}-x_{K}\|$ and
transposition-dependent invariants $\mathcal{K}\left[ \varphi
_{K},\varphi _{L}\right]$ also restrict the spatial symmetry to
O$(n,\mathbb{R})$. But it is well-known that particularly
interesting models and successful analytical procedures appear
when the group of dynamical symmetries (symmetries of Lagrangian)
coincides with the kinematical group, or at least, when it is as
large a subgroup as possible. The questions arise as to the formal
possibility and physical usefulness of affinely-invariant models.
For a single affinely-rigid body such models are in a sense
possible \cite{all04,all05}. Their physical usefulness is not yet
decided, although there are some arguments supporting it. Namely,
it is quite possible that in complex media with a complicated net
of internal interactions a single element is more sensitive to its
material surrounding than to the "true" metric tensor (produced,
according to General Relativity by the gravitational field as its
"vacuum" non-excited state). The more so such a mechanism works in
defect theory. Let us also mention the concept of effective mass
\cite{Kitt,Lev} in crystals, where the kinetic energy of electrons
is not based on the "true" metric, but on the effective tensor
produced by the material surroundings. There are nice mathematical
models of the kinetic energy of a single affine body with the
kinetic energy based on the Cauchy tensor used as a metric. There
are also some physical arguments supporting such a hypothesis
\cite{all04,all05}.

The material affine invariance may seem perhaps more natural because there exist models of
continua based on very rich material symmetry. As mentioned, this is Arnold description of the
ideal incompressible fluid. It is based on infinite-dimensional group of volume-preserving
diffeomorphism. They act on the right, i.e., as material transformations. In any case,
finite-dimensional geodetic models of small grains or suspensions with kinetic energies
materially invariant under SL$(n,\mathbb{R})$ may be considered as on over-simplified,
drastically discretized version of the Arnold model.

Obviously, apparently the Euclidean invariance of the kinetic
energy of single elements and of mutual potential elements seems
to be firmly established. But nevertheless it may be a superficial
illusion and affine invariance should be at least admitted to
consideration. When one deals with a highly condensed matter and
with very small structure elements, e.g., in the nanoscale, then
the complicated structure of interactions may result in quite
unexpected results. Namely that some hypothetic phenomenological
models based only on some symmetry guiding hints may be so
realistic as (or even more realistic) than ones based on
apparently careful structural "derivation". It is so because
"derivation" in strongly interacting structured media always
neglects a lot of factors and the collective effective phenomena
are better described by phenomenological models derived on the
basis of well-established invariance assumptions. It is so, e.g.,
in such complicated structures like atomic nuclei. In any case,
symmetry principles are then rather reliable guiding hints.

For a single affine body non-trivial potentials of internal degrees of freedom are never
affinely invariant. Only constant functions on GL$(n,\mathbb{R})$ may be so. The same is true
for one-particle external potentials of the system. As we have just seen, for the purely
mutual interactions, the binary potentials of internal degrees of freedom admit affine
invariants as arguments. As we shall see, at least formally, the same is true for
translational degrees of freedom. We shall go back to this problem later on and concentrate
now on the kinetic energy terms.

There exist at least academically interesting kinetic energies for single affine bodies, and
obviously, for their systems as well (because, unlike potentials, the kinetic energy is
additive).

Of course, the usual d'Alembert model (\ref{6}) of $T_{\rm int}$ is isotropic in the physical
space and affinely invariant under material affine transformations. This is also explicitly
visualized by its another equivalent representations,
\begin{equation}
T_{\rm tr}^{\rm is-af}=\frac{M}{2}\textrm{Tr}\left( vv^{T}\right) =\frac{M}{2}
v^{T}v=\frac{M}{2}\widehat{v}^{T}G\left[ \varphi \right] \widehat{v}=\frac{M
}{2}\textrm{Tr}\left( \widehat{v}\widehat{v}^{T}G\left[ \varphi \right] \right). \label{63}
\end{equation}
And the same holds for any of the structural element, i.e., for any $T_{A}$. Obviously, $v$,
$\widehat{v}$, $G$ must be then replaced by the corresponding $v_{A}$, $\widehat{v}_{A}$,
$G_{A}=G[\varphi_{A}]$.

The labels "tr", "is", "af" in (\ref{63}) refer respectively to "translational part",
"isometry invariant in the physical space" (thus, written on the left-hand-side:
left-invariant), and "affinely invariant in the material space of a single element" (thus,
written on the right-hand-side: right-invariant).

From the purely academic point of view one can also think about $T^{\rm af-is}_{\rm tr}$, the
model affinely invariant in the physical space and isometry invariant in the material space
(respectively, left and right invariance in GL$(n,\mathbb{R})$). It will have the form:
\begin{equation}
T_{\rm tr}^{\rm af-is}=\frac{M}{2}\widehat{v}^{T}\widehat{v}=\frac{M}{2}\textrm{Tr} \left(
\widehat{v}\widehat{v}^{T}\right) =\frac{M}{2}v^{T}C\left[ \varphi \right]
v=\frac{M}{2}\textrm{Tr}\left( vv^{T}C\left[ \varphi \right] \right). \label{64}
\end{equation}
The Cauchy deformation tensor is now used as the "metric" of the physical space.

Let us observe, there is no possibility to obtain translational kinetic energy which would be
affinely invariant in both the spatial and material sense, at least if we do not try some
extremely exotic things. The reason is that the affine group is non-simple in a very special
way and does not admit doubly-invariant and non-degenerate twice covariant tensor fields. Let
us observe that the trick with the Cauchy tensor substituted instead of the "usual" metric
tensor may be as well repeated for the internal part of (\ref{6}). This would lead to the
following expressions:
\begin{equation}
T_{\rm tr}^{\rm af-J}=\frac{1}{2}\textrm{Tr}\left( C\left[ \varphi \right] \xi J\xi
^{T}\right) =\frac{1}{2}\textrm{Tr}\left( \widehat{\Omega }J\widehat{\Omega } ^{T}\right).
\label{65}
\end{equation}
Here the upper-case labels at $T_{\rm int}$ mean that it is invariant under all spatial affine
transformations and under the material group O$(n,J)\subset$ GL$(n,\mathbb{R})$ of
$J$-preserving material transformations ("material isometries" when interpreting $J$ as kind
of the "material metric tensor"). The expression (\ref{65}) becomes $T_{\rm int}^{\rm af-is}$,
i.e., materially isotropic in the usual sense when $J$ is spherical, i.e., $J=I{\rm Id}_{n}$,
where $I$ is the scalar inertial parameter and Id$_{n}$ is the unit $n\times n$ matrix. Then
we have
\begin{equation}
T_{\rm tr}^{\rm af-is}=\frac{I}{2}\textrm{Tr}\left( \widehat{\Omega }\widehat{ \Omega
}^{T}\right) =\frac{I}{2}\textrm{Tr}\left( C\left[ \varphi \right] \xi \xi ^{T}\right).
\label{66}
\end{equation}

Let us observe that the spatially affine models (\ref{64}), (\ref{65}) are
constant-coefficient quadratic forms only when expressed in terms of non-holonomic velocities
$\widehat{\Omega}$, therefore, the metric tensors underlying them are curved, essentially
Riemannian.

One can wonder what would be a possibility symmetric to (\ref{65}), i.e., affinely invariant
in the material sense. In this sense the natural modification of $T_{\rm int}$ in (\ref{6})
would be
\begin{equation}
T_{\rm tr}^{\rm H-af}=\frac{1}{2}\textrm{Tr}\left( \Omega H\Omega ^{T}\right) =
\frac{1}{2}\textrm{Tr}\left( \xi H\left[ \varphi \right] \xi ^{T}\right), \label{67}
\end{equation}
where $H$ is a fixed positively definite matrix, and $H[\varphi]$ is its following
$\varphi^{-1}$-transform:
\begin{equation}
H\left[ \varphi \right] =\varphi ^{-1}H\varphi ^{-1T}. \label{68}
\end{equation}
So, from the point of view of the usual d'Alembert theory $H[\varphi]$ is a strange
$\varphi$-dependent although co-moving inertial tensor, whereas its spatial representation is
constant. In the usual d'Alembert affine dynamics the co-moving $J$ is constant, whereas its
spatial representation $J[\varphi]$ is configuration-dependent, thus, also time-dependent:
\begin{equation}
J\left[ \varphi \right] =\varphi J\varphi ^{T}.  \label{69}
\end{equation}

The above affine models of $T_{\rm int}$ are very peculiar in the sense that the inertial
terms in (\ref{65}), (\ref{67}) are factorized into tensor products. (The factors additional
to $H$, $J$ are the physical and material metric tensors. They are apparently absent because
we use Cartesian coordinates which analytically reduce the metric tensors to identity
matrices.) The most general $T_{\rm int}$ invariant under spatial and material affine
transformations are respectively given by
\begin{eqnarray}
T_{\rm int}^{\rm l-af}&=&\frac{1}{2}\mathcal{L}^{B}{}_{A}{}^{D}{}_{C}{}\widehat{ \Omega
}^{A}{}_{B}{}\widehat{\Omega }^{C}{}_{D},\label{70}\\
T_{\rm int}^{\rm r-af}&=&\frac{1}{2}\mathcal{R}^{j}{}_{i}{}^{l}{}_{k}{}\Omega
^{i}{}_{j}{}\Omega ^{k}{}_{l},\label{71}
\end{eqnarray}
where $\mathcal{L}$, $\mathcal{R}$ are constant and symmetric in their bi-indices (under
exchanging mutually the first and second pairs of indices).

The only model of $T_{\rm int}$ invariant simultaneously under spatial and material affine
transformations (left and right regular translations in GL$(n,\mathbb{R})$) has the form:
\begin{equation}
T_{\rm int}^{\rm af-af}=\frac{A}{2}\textrm{Tr}\left( \Omega ^{2}\right) +\frac{B}{2}
\left(\textrm{Tr}\; \Omega \right) ^{2}=\frac{A}{2}\textrm{Tr}\left( \widehat{ \Omega
}^{2}\right) +\frac{B}{2}\left(\textrm{Tr}\; \widehat{\Omega }\right) ^{2}, \label{72}
\end{equation}
where $A$, $B$ are inertial constants.

This affine-affine model of $T_{\rm int}$ is never
positively-definite. One can show that this "failure" is
non-embarrassing, moreover, it may be just profitable
\cite{all04,all05}.

It was told above that the translational part $T_{\rm tr}$ is never simultaneously left
(spatial) and right (material) affinely invariant. The highest available symmetries of $T_{\rm
tr}$ are $T^{\rm is-af}_{\rm tr}$ and $T^{\rm af-is}_{\rm tr}$ (\ref{63}), (\ref{64}). This
raises the question as to the internal counterparts $T^{\rm is-af}_{\rm int}$ and $T^{\rm
af-is}_{\rm int}$. One can easily show they are given by
\begin{eqnarray}
T_{\rm int}^{\rm is-af}&=&\frac{I}{2}\textrm{Tr}\left( \Omega ^{T}\Omega \right) +T_{\rm
int}^{\rm af-af},
\label{73}\\
T_{\rm int}^{\rm af-is}&=&\frac{I}{2}\textrm{Tr}\left( \widehat{\Omega }^{T}\widehat{ \Omega
}\right) +T_{\rm int}^{\rm af-af},  \label{74}
\end{eqnarray}
where $I$ is an additional inertial constant. In the special case of the metrically-rigid
(gyroscopic) motion, the first terms in (\ref{73}), (\ref{74}) coincide. In general this is
not the case. Therefore, if gyroscopic constraints are imposed (no deformations), (\ref{73})
and (\ref{74}) coincide, and one obtains the spherical rigid body with the scalar inertial
momentum $(I-A)$.

Unlike (\ref{72}), expressions (\ref{73}), (\ref{74}) may be positively definite and they are
so in some open range of triples $(I,A,B)$. And at the same time they continue to have the
main advantages and the general structure of (\ref{72}), both on the level of theoretical
analysis and practical calculations.

Obviously, everything said above concerns directly a single affinely-defor\-mable element,
nevertheless, just as in the d'Alembert model, applies also immediately to the total system,
because the modified kinetic energies are additive. One should only use explicitly the label
$K$ referring to structural elements.

And now let us go back to the problem of potential energy. As mentioned, the external
one-particle potential $\mathcal{V}^{(1)}$ cannot be affinely invariant (only constant
functions may be so). The formula (\ref{62}) for the doubly isotropic binary potential
$\mathcal{V}^{(2)}_{KL}$ seems to suggest something similar for the dynamics of mutual
interactions. However, things are not so simple and one can try to find some modifications
towards the affine invariance, just as it was done in the case of kinetic energy models. Some
at least formally admissible suggestion may be easily formulated.

Let us fix some pair of structural elements labelled by $(K,L)$. Internal configurations
$\varphi_{K},\varphi_{L}\in$ GL$(n,\mathbb{R})$ give rise to the Cauchy tensors
$C[\varphi_{K}]$, $C[\varphi_{L}]$. In our considerations above we were faced with the idea of
using $C[\varphi]$ as a kind of spatial "metric tensor" underlying affinely-invariant kinetic
energies of single elements. Let us now introduce the objects
\[
C\left[ \varphi _{K},\varphi _{L}\right] =\frac{1}{2}\left(
C\left[ \varphi _{K}\right] +C\left[ \varphi _{L}\right] \right).
\]
It is symmetric in the labels $K$, $L$ and positively definite. This motivates the temptation
to use it as a "metric tensor" underlying some modified "distance" between $x_{K}$ and
$x_{L}$, namely,
\begin{eqnarray}
\mathcal{D}\left[ x_{K},\varphi _{K};x_{L},\varphi _{L}\right] &=&\sqrt{ \left(
x_{K}-x_{L}\right) ^{T}C\left[ \varphi
_{K},\varphi _{L}\right]\left( x_{K}-x_{L}\right) }  \label{75} \\
&=&\sqrt{\textrm{Tr}\left( C\left[ \varphi _{K},\varphi
_{L}\right] \left( x_{K}-x_{L}\right) \left( x_{K}-x_{L}\right)
^{T}\right) }. \nonumber
\end{eqnarray}
Obviously, the transformation rule (\ref{15}) implies that the above prescription is invariant
under the spatial action of GL$(n,\mathbb{R})$:
\begin{equation}
\mathcal{D}\left[ Ax_{K},A\varphi _{K};Ax_{L},A\varphi _{L}\right] =\mathcal{ D}\left[
x_{K},\varphi _{K};x_{L},\varphi _{L}\right] \label{76}
\end{equation}
for any $A\in$ GL$(n,\mathbb{R})$. This is a rather curious affinely-invariant "distance". And
now we can modify (\ref{62}) by introducing to it this new distance-like argument in addition
to the usual one:
\[
\left\| x_{K}-x_{L}\right\| =\sqrt{\left( x_{K}-x_{L}\right) ^{T}\left( x_{K}-x_{L}\right)
}=\sqrt{\textrm{Tr}\left( \left( x_{K}-x_{L}\right) \left( x_{K}-x_{L}\right) ^{T}\right) }.
\]
So, finally, instead of (\ref{62}) we have
\begin{eqnarray}
&&\mathcal{V}_{KL}^{(2)}\left( x_{K},\varphi _{K};x_{L},\varphi _{L}\right)=\label{77}\\
&&=f_{KL}\left( \left\| x_{K}-x_{L}\right\| ,\mathcal{D}\left[ x_{K},\varphi
_{K};x_{L},\varphi _{L}\right] ,\mathcal{K}\left[ \varphi _{K},\varphi _{L} \right]
,\mathcal{M}\left[ \varphi _{K},\varphi _{L}\right] \right),\nonumber
\end{eqnarray}
where in realistic situations all $f_{KL}$ coincide with some fixed $f$. It is seen that
$\mathcal{V}_{KL}^{(2)}$ depends on its configuration arguments through the system of four
scalar quantities. Two of them, namely, $\left\| x_{K}-x_{L}\right\|$ and $\mathcal{K}\left[
\varphi _{K},\varphi _{L} \right]$ are invariants of the rotation group O$(n,\mathbb{R})$. The
remaining two, $\mathcal{D}\left[ x_{K},\varphi _{K};x_{L},\varphi _{L}\right]$ and
$\mathcal{M}\left[ \varphi _{K},\varphi _{L}\right]$, are invariant under the total linear
group GL$(n,\mathbb{R})$. One can expect that the dependence of $\mathcal{V}^{(2)}$ on the
latter two scalars is a highly symmetric, affine background of mutual interactions between
constituents of the body. Further on, this high affine symmetry is broken and reduced to the
orthogonal one O$(n,\mathbb{R})$ by the arguments $\left\| x_{K}-x_{L}\right\|$,
$\mathcal{K}\left[ \varphi _{K},\varphi _{L} \right]$. This may happen in such a way that
$\mathcal{V}^{(2)}$ is a sum of some purely affine term dependent only on $\mathcal{D}$,
$\mathcal{M}$ and on an appropriate symmetry-restricting metrical term built of $\|\cdot\|$
and $\mathcal{K}$.

It is a very interesting question whether the binary purely affine models
\begin{equation}
\mathcal{V}_{KL}^{(2)\rm af}=f_{KL}\left( \mathcal{D},\mathcal{M}\right) \label{78}
\end{equation}
may be realistic. The question was not yet touched seriously.
Nevertheless, some limitations of applicability of the binary
affine paradigm seem to be obvious. In our earlier papers
\cite{all04,all05} we discussed dynamical models of a single
affinely-rigid body, in particular, the purely geodetic models,
i.e., ones without potentials. Lagrangian coincides then with the
kinetic energies (metric tensors on GL$(n,\mathbb{R})$) given by
(\ref{70}) or (\ref{71}), and first of all, by their special cases
like (\ref{72}), (\ref{73}), (\ref{74}). It turns out that for
incompressible affine bodies, when the configuration space of
internal motion is restricted to SL$(n,\mathbb{R})$, the purely
geodetic affine models predict the existence of an open family of
bounded (oscillatory) trajectories within the general solution.
However, on the non-restricted GL$(n,\mathbb{R})$, when the volume
changes are admitted, geodetic affine models predict the
non-restricted dilatational motion, i.e., unlimited expansion or
contraction. This is an evidently non-physical feature of these
models. Therefore, at least some dilatations-stabilizing potential
$\mathcal{V}_{\rm dil}(\det\varphi)$ must be assumed. When we deal
with systems of affine bodies, then it is clear that for an
appropriate choice of $f$ in (\ref{78}) the relative volumes
\begin{equation}
\det \varphi _{L}/\det \varphi _{K}=\det \left( \varphi
_{K}{}^{-1}\varphi _{L}\right) =\det \Gamma \left[ \varphi
_{K},\varphi _{L} \right]   \label{79}
\end{equation}
are stabilized in the sense of performing bounded motions. However, no binary potential may
stabilize the single volumes $\det\varphi_{K}$ themselves. Their time evolution will be
non-bounded although the ratios (\ref{79}) are bounded functions of time. To prevent this one
should introduce some one-body potential term stabilizing (making bounded) the over-all
dilatational behaviour,
\begin{equation}
\mathcal{V}_{\rm dil}^{(1)}\left( \ldots, \varphi _{A},\ldots \right)
=\sum^{N}_{K=1}\mathcal{V}_{\rm dil}^{(1)}{}_{K}\left( \det \varphi _{K}\right). \label{80}
\end{equation}
It is reasonable to assume that all $\mathcal{V}_{\rm
dil}^{(1)}{}_{K}$ are identical (when the body consists of
identical elements),
\begin{equation}
\mathcal{V}_{\rm dil}^{(1)}\left( \ldots, \varphi _{A},\ldots \right) =\sum^{N}_{K=1}f\left(
\det \varphi _{K}\right).\label{81}
\end{equation}
When $\mathcal{V}_{KL}^{(2)}$ depend on their arguments in a proper way, so that
$\det\Gamma[\varphi_{K},\varphi_{L}]$ are bounded functions of time, then in principle it
would be sufficient to use $\mathcal{V}_{\rm dil}^{(1)}$ depending on $\det\varphi_{A}$ for
some fixed label $A$ only. If such $\mathcal{V}^{(1)}(\det\varphi_{A})$ stabilizes
$\det\varphi_{A}$, then automatically all volumes $\det\varphi_{K}$ will be stabilized by
$\mathcal{V}^{(2)}$. But of course such a choice of the shape of $\mathcal{V}$ would not be
either aesthetic or reasonable.

\section{General quantization ideas}

There is a direct logical chain from the atomic and molecular
structure to macroscopic properties, constitutive laws and
material engineering. The point is particularly delicate on the
nano-level, where one is dealing with a very peculiar convolution
of quantum and classical concepts. In any case, quantization is
necessary then. Also some quasi-classical and correspondence
problems are very relevant for these phenomena.

The first step towards quantization is the classical canonical
formalism \cite{Abr,Arn,Mac}. One should start from Legendre
transformations which for potential systems with Lagrangians
$L=T-\mathcal{V}(\cdots;x_{K},\varphi_{K};\cdots)$ are given by
\begin{equation}
p^{K}{}_{i}=\frac{\partial L}{\partial v^{i}{}_{K}}=\frac{\partial
T}{
\partial v^{i}{}_{K}},\quad \pi ^{Ka}{}_{i}=\frac{\partial L}{\partial \xi
_{K}{}^{i}{}_{a}}=\frac{\partial T}{\partial \xi _{K}{}^{i}{}_{a}}
\label{82}
\end{equation}
or, alternatively,
\begin{eqnarray}
\widehat{p}^{K}{}_{a}=\frac{\partial L}{\partial
\widehat{v}^{a}{}_{K}}= \frac{\partial T}{\partial
\widehat{v}^{a}{}_{K}},&\quad& \widehat{\Sigma }
^{Ka}{}_{b}=\frac{\partial L}{\partial \widehat{\Omega
}_{K}{}^{b}{}_{a}}= \frac{\partial T}{\partial \widehat{\Omega
}_{K}{}^{b}{}_{a}},\label{83}\\
&\quad&\Sigma^{Ki}{}_{j}=\frac{\partial
L}{\partial\Omega_{K}{}^{j}{}_{i}}=\frac{\partial T}{\partial
\Omega_{K}{}^{j}{}_{i}}.\nonumber
\end{eqnarray}
Inverting these formulas and substituting them to the energy
expression
\begin{equation}\label{84}
E=v^{i}{}_{K}\frac{\partial L}{\partial v^{i}{}_{K}}+\xi
_{K}{}^{i}{}_{a} \frac{\partial L}{\partial \xi
_{K}{}^{i}{}_{a}}-L
\end{equation}
one obtains the classical Hamiltonian
\begin{equation}
H=\mathcal{T}+\mathcal{V}.\label{75a}
\end{equation}

Let us quote the resulting formulas for the geodetic (kinetic)
Hamiltonians $\mathcal{T}$. For the "usual" d'Alembert model
(\ref{6}) we obtain
\begin{equation}
\mathcal{T}^{\rm d'A}=\mathcal{T}_{\rm tr}^{\rm
d'A}+\mathcal{T}_{\rm int}^{ \rm
d'A}=\frac{1}{2M}\textrm{Tr}\left(pp^{T}\right)+\frac{1}{2}\textrm{Tr}
\left(\pi^{T}J^{-1}\pi\right).\label{76a}
\end{equation}

This is, as mentioned, the "usual" expression compatible with the
d'Alem\-bert principle. Although from some point of view it seems
the best-motivated one, in complicated systems with collective
modes and strong internal interactions some doubts and just
objections may be raised against it. Our idea here was to
concentrate on models motivated by symmetry principles, first of
all, by affine symmetry. Let us now review Legendre transforms of
affine models quoted above.

Of course, the model (\ref{63}) of translational kinetic energy
$T^{\rm is-af}_{\rm tr}$ coincides exactly with $T^{\rm d'A}_{\rm
tr}$, so we have
\begin{eqnarray}
\mathcal{T}_{\rm tr}{}^{\rm
is-af}&=&\frac{1}{2M}p^{T}p=\frac{1}{2M}\textrm{Tr}\left(
pp^{T}\right)\nonumber\\
&=&\frac{1}{2M}\widehat{p}^{T}G[\varphi]^{-1}
\widehat{p}=\frac{1}{2M}\textrm{Tr}\left(
\widehat{p}\widehat{p}^{T}G[\varphi]^{-1}\right), \label{77a}
\end{eqnarray}
just the first term of (\ref{76a}) written in a few equivalent
forms.

The corresponding expression for (\ref{64}) has the following
form:
\begin{eqnarray}
\mathcal{T}_{\rm tr}{}^{\rm
af-is}&=&\frac{1}{2M}\widehat{p}^{T}\widehat{p}=\frac{1}{
2M}\textrm{Tr}\left(\widehat{p}\widehat{p}^{T}\right)\nonumber\\
&=&\frac{1}{2M}p^{T}C[\varphi]^{-1}p=\frac{1}{2M}\textrm{Tr}\left(
pp^{T}C[\varphi]^{-1}\right).\label{78a}
\end{eqnarray}

Let us now quote the Legendre transforms of affinely-invariant
internal kinetic energies. For (\ref{65}) we obtain
\begin{equation}
\mathcal{T}_{\rm int}^{\rm af-J}=\frac{1}{2}\textrm{Tr}\left(
C[\varphi] ^{-1}\pi ^{T}J^{-1}\pi \right)
=\frac{1}{2}\textrm{Tr}\left( \widehat{\Sigma
}^{T}J^{-1}\widehat{\Sigma }\right).  \label{79a}
\end{equation}
In particular, for the isotropic inertial tensor (\ref{66}) leads
to the expression:
\begin{equation}
\mathcal{T}_{\rm int}^{\rm af-is}=\frac{1}{2I}\textrm{Tr}\left(
\widehat{\Sigma } ^{T}\widehat{\Sigma }\right)
=\frac{1}{2I}\textrm{Tr}\left( C[\varphi]^{-1}\pi ^{T}\pi \right).
\label{80a}
\end{equation}
For the model (\ref{67}) we obtain
\begin{equation}
\mathcal{T}_{\rm int}^{\rm H-af}=\frac{1}{2}\textrm{Tr}\left(
\Sigma ^{T}H^{-1}\Sigma \right)=\frac{1}{2}\textrm{Tr}\left( \pi
^{T}H\left[ \varphi \right] ^{-1}\pi \right).  \label{81a}
\end{equation}
Obviously, the Legendre transforms of the most general affine
models (\ref{70}), (\ref{71}) have the form:
\begin{eqnarray}
\mathcal{T}_{\rm int}^{\rm
l-af}&=&\frac{1}{2}\widetilde{\mathcal{L}}^{B}{}_{A}{}^{D}{}_{C}
\widehat{\Sigma }^{A}{}_{B}\widehat{\Sigma }^{C}{}_{D},\label{82a}\\
\mathcal{T}_{\rm int}^{\rm
r-af}&=&\frac{1}{2}\widetilde{\mathcal{R}}^{j}{}_{i}{}^{l}{}_{k}
\Sigma ^{i}{}_{j}\Sigma ^{k}{}_{l},\label{83a}
\end{eqnarray}
where the constant bimatrices $\widetilde{\mathcal{L}}$,
$\widetilde{\mathcal{R}}$ are reciprocal to $\mathcal{L}$,
$\mathcal{R}$ respectively.

For the most interesting models (\ref{73}), (\ref{74}), including
their special doubly-affine case (\ref{72}) we obtain respectively
\begin{eqnarray}
\mathcal{T}_{\rm int}^{\rm
is-af}&=&\frac{1}{2\widetilde{I}}\textrm{Tr}\left( \Sigma
^{T}\Sigma \right) +\frac{1}{2\widetilde{A}}\textrm{Tr}\left(
\Sigma ^{2}\right)
+\frac{1}{2\widetilde{B}}\left(\textrm{Tr} \Sigma \right) ^{2},\label{84a}\\
\mathcal{T}_{\rm int}^{\rm
af-is}&=&\frac{1}{2\widetilde{I}}\textrm{Tr}\left( \widehat{\Sigma
}^{T}\widehat{\Sigma }\right) +\frac{1}{2\widetilde{A}}
\textrm{Tr}\left( \widehat{\Sigma }^{2}\right)
+\frac{1}{2\widetilde{B}} \left(\textrm{Tr} \widehat{\Sigma
}\right)^{2},  \label{85}
\end{eqnarray}
where the constants $\widetilde{I}$, $\widetilde{A}$,
$\widetilde{B}$ are built of $I$, $A$, $B$ in the following way:
\begin{equation}
\widetilde{I}=\frac{1}{I}\left( I^{2}-A^{2}\right),\
\widetilde{A}=\frac{1}{ A}\left( A^{2}-I^{2}\right),\
\widetilde{B}=-\frac{1}{B}\left( I+A\right) \left(
I+A+nB\right).\label{86}
\end{equation}
The special affine-affine case (\ref{72}) corresponds to $I=0$,
and then obviously $1/\widetilde{I}=0$ and the first terms of
(\ref{84a}), (\ref{85}) do vanish. Of course, the second and third
terms of (\ref{84a}), (\ref{85}) are pairwise identical.

These were expressions for various models of the kinetic
Hamiltonians for single elements. Obviously, they should be
labelled by the index $K=\overline{1,N}$, and the total expression
is obtained by a summation over $K$.

Canonical formalism is very convenient and effective in analysis
of classical equations of motion. They are represented then in
terms of Poisson brackets,
\begin{equation}
\frac{dF}{dt}=\left\{ F,H\right\},  \label{87}
\end{equation}
where $F$ runs over some maximal functionally independent system
of functions. To make use of (\ref{87}) one must establish the
system of basic Poisson brackets for some geometrically
distinguished quantities. As a rule, one uses quantities like
$\Sigma$, $\widehat{\Sigma}$, because they are Hamiltonian
generators of left and right regular translations in
GL$(n,\mathbb{R})$. Therefore, their Poisson brackets are
determined by the structure constants of this group (and the
mutual Poisson brackets between $\Sigma$ and
$\widehat{\Sigma}$-quantities do vanish of course, because the
left translations commute with the right ones). Other important
Poisson brackets are those between
$\Sigma,\widehat{\Sigma}$-quantities and functions depending only
on generalized coordinates. Calculating such a bracket is
identical with affecting configuration functions by first-order
differential operators generating left and right regular
translations in GL$(n,\mathbb{R})$.

However, the main advantage of Hamiltonian methods is that they
provide a direct way towards quantization.

Let us remind that the configuration space of our $N$-body system
is given by
\[
Q^{N}\simeq Q^{N}{}_{\rm tr}\times Q^{N}{}_{\rm int}\simeq
\mathbb{R}^{nN}\times{\rm GL}\left(n,\mathbb{R}\right)^{N},
\]
i.e., configurations are arrays (\ref{5}). The manifold $Q^{N}$ is
obviously an open subset of the linear space
\[
\mathbb{R}^{nN}\times{\rm L}\left(n,\mathbb{R}\right)^{N}\simeq
\mathbb{R}^{nN}\times\mathbb{R}^{n^{2}N}\simeq\mathbb{R}^{n(n+1)N}.
\]
In any one-element configuration space $\mathbb{R}^{n}\times$
GL$(n,\mathbb{R})$ we are given two distinguished measures. One of
them is the Haar measure $\alpha$ invariant under left and right
group translations (cf. \cite{all04}). The other one is the usual
Lebesgue measure $a$ on $\mathbb{R}^{n}\times$ L$(n,\mathbb{R})$.
It is invariant under additive translations. In terms of
coordinates
\begin{eqnarray}
da\left( x,\varphi \right) &=&dx^{1}\cdots dx^{n}d\varphi
^{1}{}_{1}\cdots d\varphi ^{n}{}_{n},  \label{88}\\
d\alpha \left( x,\varphi \right) &=&\left( \det \varphi \right)
^{-n-1}da\left( x,\varphi \right)\nonumber\\
&=&\left( \det \varphi \right) ^{-n-1}dx^{1}\cdots dx^{n}d\varphi
^{1}{}_{1}\cdots d\varphi ^{n}{}_{n}. \label{89}
\end{eqnarray}
When we neglect translational motion then the Haar measure
$\lambda$ on GL$(n,\mathbb{R})$ and the Lebesgue measure $l$ on
L$(n,\mathbb{R})$ are used
\begin{eqnarray}
dl \left( \varphi \right)&=&d\varphi ^{1}{}_{1}\cdots d\varphi
^{n}{}_{n}, \label{90}\\
d\lambda \left( \varphi \right)&=&\left( \det \varphi \right)
^{-n-1}dl \left( \varphi \right) =\left( \det \varphi \right)
^{-n}d\varphi ^{1}{}_{1}\cdots d\varphi ^{n}{}_{n}.  \label{91}
\end{eqnarray}
Configuration spaces of the total $N$-element system are endowed
with the $N$-told tensor products of these measures, $a^{(N)}$,
$\alpha^{(N)}$, $l^{(N)}$, $\lambda^{(N)}$.

The quantized theory is formulated in the following Hilbert spaces:
\[
{\rm L}^{2}\left(Q^{N},\alpha^{(N)}\right), {\rm L}^{2}\left(Q^{N},a^{(N)}\right), {\rm
L}^{2}\left({\rm GL}(n,\mathbb{R})^{N},\lambda^{(N)}\right), {\rm L}^{2}\left({\rm
GL}(n,\mathbb{R})^{N},l^{(N)}\right).
\]
Their elements, i.e., wave functions, are complex probability amplitudes of finding the system
at a given classical configuration. Classical quantities depending only on configuration
variables are represented in these L$^{2}$-spaces as operators of multiplication by
real-valued functions, in particular, by coordinates like $x^{i}$, $\varphi^{i}{}_{a}$, etc.
According to the general rules of quantum mechanics all other quantities are also represented
by Hermitian or formally Hermitian (symmetric in dense domains) operators in these Hilbert
spaces. Usually some ordering problems of non-commuting operators appear then. However, in
dynamical applications, when Hamiltonian operators are constructed, one deals usually with
some special physical quantities of well-defined geometric interpretation. As a rule, they are
generators of symmetry groups underlying the problem. In our model they are just the affine
spin in both the spatial and co-moving representation, the usual metrical spin and vorticity,
etc.

Linear momentum operators in spatial and co-moving representations
are given respectively by
\begin{equation}
\mathbf{p}^{K}{}_{a}=\frac{\hbar }{i}\frac{\partial }{\partial
x^{a}{}_{K}} ,\qquad \widehat{\mathbf{p}}^{K}{}_{a}=\frac{\hbar
}{i}\varphi _{K}\mathbf{} ^{b}{}_{a}\frac{\partial }{\partial
x^{b}{}_{K}},  \label{92}
\end{equation}
where, obviously, $K=\overline{1,N}$ is the "particle" label.
These operators are formally Hermitian both in
L$^{2}\left(Q^{N},\alpha^{(N)}\right)$ and
L$^{2}\left(Q^{N},a^{(N)}\right)$. The operators
\begin{equation}
\mathbf{\Sigma }_{K}\mathbf{}^{a}{}_{b}=\frac{\hbar }{i}\varphi
_{K}\mathbf{} ^{a}{}_{c}\frac{\partial }{\partial \varphi
_{K}\mathbf{}^{b}{}_{c}},\qquad \widehat{\mathbf{\Sigma
}}_{K}\mathbf{}^{a}{}_{b}=\frac{\hbar }{i}\varphi
_{K}\mathbf{}^{c}{}_{b}\frac{\partial }{\partial \varphi
_{K}\mathbf{} ^{c}{}_{a}}\label{93}
\end{equation}
are formally Hermitian (not literally, they are unbounded as all
differential operators) in L$^{2}\left(Q^{N},\alpha^{(N)}\right)$
and in L$^{2}\left({\rm
GL}(n,\mathbb{R})^{N},\lambda^{(N)}\right)$. Therefore, when using
these Hilbert spaces we may interpret (\ref{93}) as operators of
affine spin respectively in the spatial and co-moving
representations.

Just as in classical theory, $\mathbf{p}^{K}{}_{a}$ are infinitesimal generators of
translations of the $K$-th constituent. Similarly, $\mathbf{\Sigma }_{K}\mathbf{}^{a}{}_{b}$
generate spatial affine transformations (rotations and homogeneous deformations) of internal
degrees of freedom of the $K$-th "molecule". $\widehat{\mathbf{\Sigma
}}_{K}\mathbf{}^{a}{}_{b}$ generate material affine transformations of the $K$-th element.
Namely, let us consider the operators
\begin{eqnarray}
\mathbf{V}_{K}\left( y\right)&:=&\exp \left( \frac{i}{\hbar }y^{a}\mathbf{p}
^{K}{}_{a}\right),\qquad y\in \mathbb{R}^{n}, \label{94} \\
\mathbf{L}_{K}\left( z\right)&:=&\exp \left( \frac{i}{\hbar }z^{b}{}_{a} \mathbf{\Sigma
}_{K}\mathbf{}^{a}{}_{b}\right),\qquad z\in{\rm L}\left( n,\mathbb{R}\right),  \label{95}
\end{eqnarray}
where the operator exponent is meant in the usual power-series sense. If this series
convergent in the action on some function $\Psi:Q^{N}\rightarrow \mathbb{C}$, then
\begin{eqnarray}
&&\left( \mathbf{V}_{K}\left( y\right) \Psi \right) \left( \ldots ,x_{A},\ldots ;\ldots
,\varphi_{B},\ldots \right)=\label{96}\\
&&\hspace{3cm}=\Psi \left( \ldots ,x_{A}+y\delta _{AK},\ldots ;\ldots ,\varphi _{B},\ldots
\right),\nonumber\\
&&\left( \mathbf{L}_{K}\left( z\right) \Psi \right) \left( \ldots ,x_{A},\ldots ;\ldots
,\varphi _{B},\ldots \right)=\label{97}\\
&&\hspace{3cm}=\Psi \left( \ldots ,x_{A}\ldots ;\ldots ,\exp \left( z\delta _{KB}\right)
\varphi _{B},\ldots \right).\nonumber
\end{eqnarray}
Similar statements may be formulated about the co-moving objects, e.g., defining
\begin{equation}
\mathbf{R}_{K}\left( z\right):=\exp \left( \frac{i}{\hbar }z^{b}{}_{a} \widehat{\mathbf{\Sigma
}}_{K}{}^{a}{}_{b}\right),\qquad z\in{\rm L}\left( n,\mathbb{R}\right), \label{98}
\end{equation}
we obtain that
\begin{eqnarray}
&&\left( \mathbf{R}_{K}\left( z\right) \Psi \right) \left( \ldots ,x_{A},\ldots ;\ldots
,\varphi _{B,}\ldots \right)=\label{99}\\
&&\hspace{3cm}=\left( \ldots ,x_{A},\ldots ;\ldots ,\varphi _{B}\exp \left( z\delta
_{BK}\right), \ldots \right).\nonumber
\end{eqnarray}
One can act separately on all arguments, nevertheless, the special geometric role is played by
transformations acting in the same way on all arguments, e.g.,
\begin{eqnarray}
\mathbf{V}\left( y\right)&=&\mathbf{V}_{1}\left( y\right) \cdots
\mathbf{V} _{N}\left( y\right),  \label{100}\\
\mathbf{L}\left( z\right)&=&\mathbf{L}_{1}\left( z\right) \cdots
\mathbf{L} _{N}\left( z\right),  \label{101}\\
\mathbf{R}\left( z\right)&=&\mathbf{R}_{1}\left( z\right) \cdots
\mathbf{R}_{N}\left(z\right).\label{102}
\end{eqnarray}
Their generators are respectively identical with the total linear momentum and the total
affine spin in the spatial and co-moving representations,
\begin{equation}
\mathbf{p}_{a}=\sum^{N}_{K=1} \mathbf{p}^{K}{}_{a},\qquad \mathbf{\Sigma
}^{a}\mathbf{}_{b}=\sum^{N}_{K=1}\mathbf{ \Sigma }_{K}{}^{a}\mathbf{}_{b},\qquad
\widehat{\mathbf{\Sigma }}^{a}\mathbf{} _{b}=\sum^{N}_{K=1}\widehat{\mathbf{\Sigma
}}_{K}{}^{a} \mathbf{}_{b}.  \label{103}
\end{equation}
Obviously,
\begin{eqnarray}
\mathbf{V}\left( y\right)&=&\exp \left( \frac{i}{\hbar
}y^{a}\mathbf{p} _{a}\right),   \label{104}\\
\mathbf{L}\left( z\right)&=&\exp \left( \frac{i}{\hbar
}z^{b}{}_{a}\mathbf{ \Sigma }^{a}{}_{b}\right),   \label{105}\\
\mathbf{R}\left( z\right)&=&\exp \left( \frac{i}{\hbar }z^{b}{}_{a}\widehat{ \mathbf{\Sigma
}}^{a}{}_{b}\right).\label{106}
\end{eqnarray}
Obviously, all exponential operators quoted here are unitary in
L$^{2}\left(Q^{N},\alpha^{(N)}\right)$ or L$^{2}\left({\rm
GL}(n,\mathbb{R})^{N},\lambda^{(N)}\right)$. However, $\mathbf{L}_{K}$, $\mathbf{R}_{K}$ are
not unitary in L$^{2}\left(Q^{N},a^{(N)}\right)$ and L$^{2}\left({\rm
GL}(n,\mathbb{R})^{N},l^{(N)}\right)$. The reason is that the measures $a$, $l$ are not
invariant under group translations. Also, when working in L$^{2}\left(Q^{N},a^{(N)}\right)$
and L$^{2}\left({\rm GL}(n,\mathbb{R})^{N},l^{(N)}\right)$, that is admissible, one must
modify the definition of the above unitary operators (introducing some multipliers). The
generators (\ref{93}) are not formally Hermitian and to become such they must be modified by
some additive corrections:
\begin{equation}
^{\prime}\mathbf{\Sigma }_{K}{}^{a}{}_{b}:=\mathbf{\Sigma }_{K}{}^{a}{}_{b}+ \frac{\hbar
n}{2i}\delta ^{a}{}_{b},\quad ^{\prime}\widehat{\mathbf{\Sigma }}
_{K}{}^{a}{}_{b}:=\widehat{\mathbf{\Sigma }}_{K}{}^{a}{}_{b}+\frac{\hbar n}{ 2i}\delta
^{a}{}_{b}.\label{107}
\end{equation}
Let us also quote the formally Hermitian operators
\begin{equation}
\mathbf{J}_{K}{}^{a}{}_{b}=\mathbf{x}_{K}\mathbf{}^{a}\mathbf{p}^{K}{}_{b}+ \mathbf{\Sigma
}_{K}{}^{a}{}_{b}=\mathbf{\Lambda }_{K}{}^{a}{}_{b}+\mathbf{ \Sigma
}_{K}{}^{a}{}_{b},\label{108}
\end{equation}
which generate affine transformations acting both on translational and internal degrees of
freedom of the $K$-th constituents. $\mathbf{\Lambda }_{K}$ and $\mathbf{ \Sigma }_{K}$ are
respectively the translational (orbital) and internal parts. One can also introduce the total
quantities
\begin{equation}
\mathbf{J}^{a}{}_{b}=\mathbf{\Lambda }^{a}{}_{b}+\mathbf{\Sigma }^{a}{}_{b} \label{109}
\end{equation}
obtained by the $K$-summation.

In analogy to (\ref{107}), we have that
\begin{equation}
^{\prime}\mathbf{\Lambda }_{K}{}^{a}{}_{b}:=\mathbf{\Lambda }_{K}{}^{a}{}_{b}+
\frac{\hbar}{2i}\delta ^{a}{}_{b}.\label{110}
\end{equation}
Let us observe that for the total quantities built of (\ref{107}) we have
\begin{equation}
^{\prime}\mathbf{\Sigma }^{a}{}_{b}=\mathbf{\Sigma }^{a}{}_{b}+\frac{\hbar nN}{ 2i}\delta
^{a}{}_{b},\quad {}^{\prime} \widehat{\mathbf{\Sigma }}^{a}{}_{b}= \widehat{\mathbf{\Sigma
}}^{a}{}_{b}+\frac{\hbar nN}{2i}\delta ^{a}{}_{b}.\label{111}
\end{equation}
and similarly
\begin{equation}
^{\prime}\mathbf{\Lambda }^{a}{}_{b}=\mathbf{\Lambda }^{a}{}_{b}+\frac{\hbar N }{2i}\delta
^{a}{}_{b}.\label{112}
\end{equation}

After quantization the canonical momenta $\pi^{Ka}{}_{i}$ conjugate to
$\varphi_{K}{}^{i}{}_{a}$ become operators:
\begin{equation}
\mathbf{p}^{Ka}{}_{j}:=\frac{\hbar }{i}\frac{\partial }{\partial
\varphi _{K}{}^{j}{}_{a}}.  \label{113}
\end{equation}
They are formally Hermitian in L$^{2}\left(Q^{N},a^{(N)}\right)$, L$^{2}\left({\rm
GL}(n,\mathbb{R})^{N},l^{(N)}\right)$, but not in L$^{2}\left(Q^{N},\alpha^{(N)}\right)$,
L$^{2}\left({\rm GL}(n,\mathbb{R})^{N},\lambda^{(N)}\right)$, so now the situation is quite
opposite to the previous one.

Hamilton operator has the following form:
\begin{equation}
\mathbf{H}=\mathbf{T}+\mathcal{V},\label{114}
\end{equation}
where $\mathbf{T}$ is the kinetic energy operator and $\mathcal{V}$ is the potential term;
usually one does not distinguish graphically between the function $\mathcal{V}$ and the
operator $\mathbf{V}$ which multiplies the wave function $\Psi$ by $\mathcal{V}$,
\begin{equation}
\left( \mathbf{V}\Psi \right) \left( x,\varphi \right):=\mathcal{V}\left( x,\varphi \right)
\Psi \left( x,\varphi \right).  \label{115}
\end{equation}
In usual structural problems quantum dynamics reduces to the stationary Schr\"o-dinger
equation, i.e., to the energy following eigenproblem:
\begin{equation}
\mathbf{H}\Psi =E\Psi.\label{116}
\end{equation}
And finally, we see that constructing the kinetic energy operator $\mathbf{T}$ is a crucial
step of the quantization procedure. And this may be done just on the basis of the above
classical expressions for kinetic energies in terms of canonical phase-space variables
(\ref{76}), (\ref{77}), (\ref{78}), (\ref{79}), (\ref{80}), (\ref{81}), (\ref{82}),
(\ref{83}), (\ref{84}), (\ref{75a}). Simply one should algebraically substitute the above
operators $\mathbf{p}^{K}{}_{i}$, $\mathbf{\widehat{p}}^{K}{}_{a}$, $\mathbf{p}^{Ka}{}_{j}$,
$\mathbf{\Sigma}^{Ka}{}_{b}$, $\mathbf{\widehat{\Sigma}}^{Ka}{}_{b}$ instead of the
corresponding classical expressions. Because of the geometric meaning of these quantities as
generators of natural transformation groups, there is no problem of ordering of operators. The
only point one should be careful with is just the one concerning the measures used in
configuration spaces when defining the L$^{2}$-Hilbert spaces. In affine models the measures
$\alpha$, $\lambda$ are more natural and then one uses the purely differential operators
(\ref{92}), (\ref{93}). In Hilbert spaces based on $a$, $l$ we would have to use the modified
expressions (\ref{107}). And conversely, in non-affine d'Alembert models the Lebesgue measures
$a$, $l$ are more natural, because they enable one to use the purely differential operators
(\ref{113}) without any algebraic correction.

The above geometric approach is very convenient. If we tried to calculate the kinetic energy
operators as
\[
\mathbf{T}=-\frac{\hbar^{2}}{2}\triangle,
\]
where $\triangle$ is the d'Alembert operator based on the metric tensor underlying the
classical expression $T$, the calculations would be hopeless and the result completely
obscure, non-useful.

When solving any particular problem one must use coordinates. For our purposes the most
convenient choice is that based on the polar and two-polar decompositions:
\[
\varphi=UA=BU=LDR^{-1}=LDR^{T},
\]
where $\varphi\in$ GL$^{+}(n,\mathbb{R})$, $U,L,R\in$ SO$(n,\mathbb{R})$ (orthogonal), $A$,
$B=UAU^{-1}$ are symmetric and positively definite, and $D$ is diagonal and positive.

Green and Cauchy deformation tensors are then expressed as follows:
\[
G=A^{2}=RD^{2}R^{T},\qquad C=B^{-2}=LD^{-2}L^{-1}.
\]
It is convenient to denote
\[
D_{aa}=Q^{a}=\exp \left( q^{a}\right).
\]
The quantities $Q^{a}$, $q^{a}$ offer another convenient choices of basic deformation
invariants. The Haar and Lebesgue measure $\lambda$, $l$ are then expressed as
\begin{eqnarray}
d\lambda \left( L,D,R\right)&=&\prod_{i\neq j}\left| {\rm sh} \left( q^{i}-q^{j}\right)
\right| dq^{1}\ldots dq^{n}d\mu \left( L\right)
d\mu \left( R\right),\nonumber\\
dl\left( L,D,R\right)&=&\prod_{i\neq j}\left( Q^{i}-Q^{j}\right) \left( Q^{i}+Q^{j}\right)
dQ^{1}\ldots dQ^{n}d\mu \left( L\right) d\mu \left( R\right),\nonumber
\end{eqnarray}
where $\mu$ is the Haar measure on SO$(n,\mathbb{R})$. Performing some partial integrations
one can obtain from $\overline{\Psi}\Psi$ probability distributions for some quantities like
deformation invariants, orientations of the main axes of Green and Cauchy deformation tensors
and also probability distributions for particular values of Green and Cauchy deformation
tensors.

\section*{Acknowledgements}

The author is greatly indebted to professors Gianfranco Capriz and
Paolo Mariano for stimulating discussions in Pisa, Warsaw, and
Berlin.

\end{document}